\pdfoutput=1

\documentclass[11pt,twoside,a4paper,cmspaper,final,collab]{cms-tdr}
\def\svnVersion{379199}\def\cmsCernNoTag{CERN-EP/2016-202}\def\cmsCernDate{\today}\def\cmsMessage{Published in the Journal of High Energy Physics as \href{http://dx.doi.org/10.1007/JHEP12(2016)123}{\doi{10.1007/JHEP12(2016)123.}}}

\begin{document}\cmsNoteHeader{TOP-15-014}

\hyphenation{had-ron-i-za-tion}
\hyphenation{cal-or-i-me-ter}
\hyphenation{de-vices}
\RCS$Revision: 374968 $
\RCS$HeadURL: svn+ssh://svn.cern.ch/reps/tdr2/papers/TOP-15-014/trunk/TOP-15-014.tex $
\RCS$Id: TOP-15-014.tex 374968 2016-11-27 21:35:52Z alverson $

\cmsNoteHeader{TOP-15-014}
\title{Measurement of the mass of the top quark in decays with a \PJGy meson in pp collisions at 8\TeV}

\date{\today}

\abstract{
  A first measurement of the top quark mass in the decay channel $\PQt\to(\PW\to\ell\nu)\,(\PQb\to\PJGy+\mathrm{X}\to\PGmp\PGmm+\mathrm{X})$ is presented. The analysis uses events selected from the proton-proton collisions recorded by the CMS detector at the LHC at a center-of-mass energy of 8\TeV. The data correspond to an integrated luminosity of 19.7\fbinv, with 666 \ttbar and single top quark candidate events containing a reconstructed \PJGy candidate decaying into an oppositely-charged muon pair. The mass of the ($\PJGy+\ell$) system, where $\ell$ is an electron or a muon from W boson decay, is used to extract a top quark mass of $173.5 \pm 3.0\stat\pm 0.9\syst\GeV$.
}

\hypersetup{%
pdfauthor={CMS Collaboration},%
pdftitle={Measurement of the mass of the top quark in decays with a J/psi meson in pp collisions at 8 TeV},%
pdfsubject={CMS},%
pdfkeywords={CMS, physics, top quark mass}}

\maketitle
\section{Introduction}
\label{sec:introduction}

The top quark is the most massive particle in the standard model (SM), with the largest Yukawa coupling to the Higgs boson. The mass of the top quark ($m_\PQt$) is a fundamental parameter of the SM, playing a key role in radiative electroweak corrections~\cite{Baak:s100052,Buras:2012ru} and likely in the mechanism of electroweak symmetry breaking~\cite{Degrassi:082012098}. Therefore, a precise determination of $m_\PQt$ is essential for a better understanding of the SM.

Since the first observation of the top quark~\cite{Abe:1995hr,Abachi:1995iq}, measurements of its mass have relied on the reconstruction of its decay products. These measurements are currently dominated by systematic uncertainties, related to the b-jet energy scale and the modeling of soft quantum chromodynamics (QCD) effects such as b quark hadronization and the underlying event~\cite{Aad:2015nba,Khachatryan:2015hba}. Currently, the most precise measurement of $m_\PQt$, $172.44\pm0.13\stat\pm0.47\syst\GeV$, is from the combination of measurements at 7 and 8\TeV by the CMS experiment~\cite{Khachatryan:2015hba}.

\begin{figure}[htb]
\centering
\includegraphics[width=0.8\textwidth]{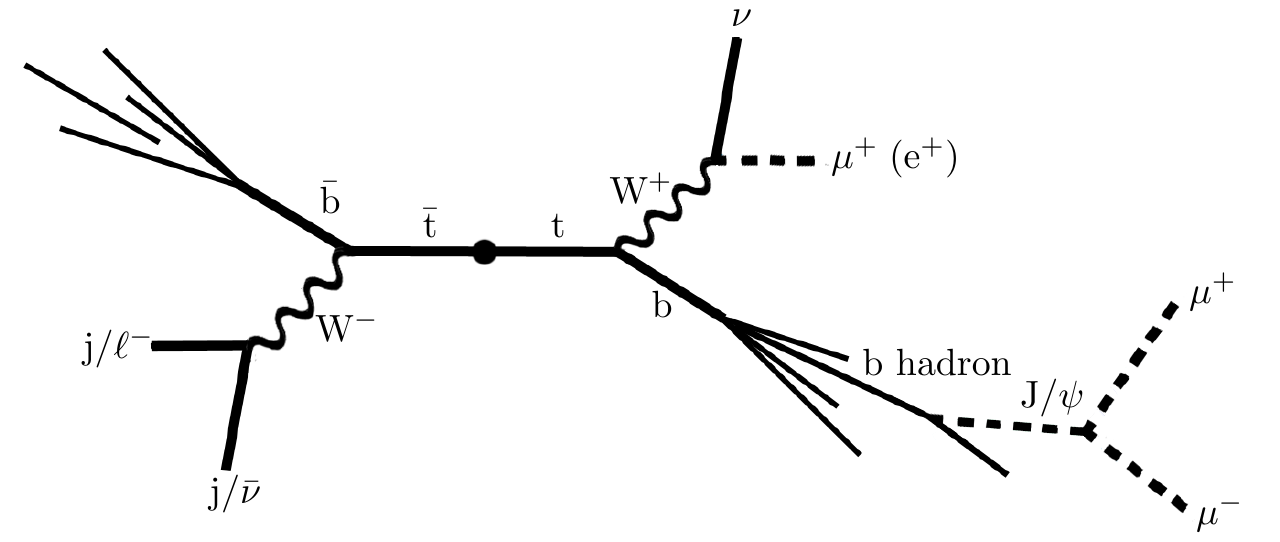}
\caption{Pictorial view of the \PJGy meson produced in a \ttbar system. The kinematic properties of the particles represented with dashed lines are used to infer $m_\PQt$.}
\label{fig:channel}
\end{figure}

In this paper, a measurement is presented of $m_\PQt$ from partial reconstruction of top quarks in leptonic final states that contain a \PJGy meson from a b hadron decay.
Both top quark-antiquark pair (\ttbar) and single top quark production are considered to be signal in this study.
The decay mode of interest is ${\PQt\to(\PW\to\ell\nu)\,(\PQb\to\PJGy+\mathrm{X}\to\PGmp\PGmm+\mathrm{X})}$ and is shown (for \ttbar production) in Fig.~\ref{fig:channel}. Here and everywhere, the charge conjugation is implicit.
As suggested in Ref.~\cite{Markytan:290808} and refined in Ref.~\cite{Kharchilava:1999yj}, the value of $m_\PQt$ is determined through its correlation with the mass of the $\PJGy+\ell$ system, where $\ell$ is either an electron or muon produced in the decay of the accompanying W boson (either directly or via a $\tau$ lepton) in the same top quark decay.
The branching fraction is expected to be $(1.5\pm0.1)\times10^{-4}$, but the presence of three leptons in the final state, two of which originate from the \PJGy meson decay, provides a nearly background-free sample of events.

This measurement is based on data collected in \Pp\Pp\ collisions at a center-of-mass energy of $8\TeV$ with the CMS detector at the CERN LHC. Simulated events generated at different top quark masses are used to calibrate the method and evaluate its performance, as well as to estimate systematic uncertainties.
The main advantage of this analysis lies in the determination of $m_\PQt$ using only leptons. In this way, the dependence of the measurement on several dominant systematic uncertainties linked to initial- and final-state radiation, jet reconstruction and b tagging techniques, is considerably reduced. The drawback is the expected sensitivity to the modeling of b quark fragmentation, and the limited number of events in the selected sample on account of the small branching fraction.

\section{Experimental setup}
\label{sec:experiment}

\subsection{The CMS detector}
\label{subsec:experiment:detector}

The central feature of the CMS apparatus is a superconducting solenoid of 6\unit{m} internal diameter, providing a magnetic field of 3.8\unit{T}.
Within the solenoid volume are a silicon pixel and strip tracker, a lead tungstate crystal electromagnetic calorimeter (ECAL), and a brass and scintillator hadron calorimeter (HCAL), each composed of a barrel and two endcap sections.
Forward calorimeters extend the pseudorapidity ($\eta$) coverage provided by the barrel and endcap detectors.
Muons are measured in gas-ionization detectors embedded in the steel flux-return yoke outside the solenoid.
The tracker has a track-finding efficiency of more than 99\% for muons with transverse momentum $\pt>1\GeV$ and $\abs{\eta}<2.5$.
The ECAL is a fine-grained calorimeter with quasi-projective geometry, and consists of a barrel region of $\abs{\eta}<1.48$ and two endcaps that extend up to $\abs{\eta}$ of $3.0$.
The HCAL barrel and endcaps similarly cover the region $\abs{\eta}<3.0$.
Muons are measured in the $\abs{\eta}<2.4$ range, with detection planes made using three technologies: drift tubes, cathode strip chambers, and resistive-plate chambers. Matching muons to tracks measured in the silicon tracker results in a relative \pt resolution for muons with $20<\pt<100\GeV$ of 1--2\% in the barrel and better than 6\% in the endcaps.
The \pt resolution in the barrel is better than 10\% for muons with \pt up to 1\TeV~\cite{Chatrchyan:2012xi}.
A more detailed description of the CMS detector, together with a definition of the coordinate systems and kinematic variables, can be found in Ref.~\cite{Chatrchyan:2008zzk}.

\subsection{Data and simulation}
\label{subsec:experiment:datasim}

This measurement is performed using the data recorded by the CMS detector at $\sqrt{s}=8\TeV$, corresponding to an integrated luminosity of $19.7\pm0.5\fbinv$~\cite{PAS-LUM-13-001}. Events are required to pass a single-muon (single-electron) trigger with a minimum muon (electron) \pt of $24\,(27)\GeV$. The method used to extract $m_\PQt$ has been developed and optimized using simulated events, without accessing the final data.

We use simulated events to develop the analysis method and estimate its performance.
The \ttbar, W$+$jets, and Z$+$jets processes are generated with the leading-order (LO) {\MADGRAPH}~\cite{Alwall:2014hca} generator (v5.1.3.30) matched to LO {\PYTHIA~6}~\cite{Sjostrand:2006za} (v6.426) for parton showering and fragmentation.
The $\tau$ lepton decays are simulated with the {\TAUOLA}~\cite{Davidson:2010rw} program (v27.121.5).
The LO CTEQ6L1~\cite{Pumplin:2002vw} parton distribution function (PDF) set and the Z2* underlying event tune are used in the generation. The most recent {\PYTHIA} Z2* tune is derived from the Z1 tune~\cite{Field:2010bc}, which uses the CTEQ5L parton distribution set, whereas Z2* adopts CTEQ6L.
Matrix elements describing up to three partons in addition to the \ttbar pair are included in the {\MADGRAPH} generator, and the MLM prescription~\cite{Mangano:2006rw} is used for matching of matrix-element jets to parton showers.
The Lund string model~\cite{PR97.1983.31} is used for the simulation of the hadronization, and to determine the fraction of the quark energy carried by unstable hadrons. For heavy quarks, the Lund symmetric fragmentation function is modified according to the Bowler space-time picture of string evolution~\cite{Bow81}. Assuming fragmentation universality~\cite{JMP.1.1724268,PR.133.B1549}, the values of the parameters of the fragmentation function obtained from fits to the LEP data~\cite{Sjo89} are used, without assigning a systematic uncertainty associated to the universality assumption.
The single top quark $t$-channel, $s$-channel, and tW processes are simulated with the next-to-leading-order (NLO) {\MADGRAPH}~\cite{Alioli:2009je,Re:2010bp} generator (v1.0, r1380) with the CTEQ6M PDF set. Diboson WW, WZ, and ZZ processes are generated with {\PYTHIA~6} (v6.426).

The simulated processes are normalized to their theoretical cross sections. Except for single top quark processes, the higher-order calculation is used, and associated systematic uncertainties are discussed in Sec.~\ref{subsec:systematics:th}. The \ttbar cross section is computed at next-to-next-to-leading-order (NNLO)~\cite{Czakon:2013goa}, while single top quark processes are computed at approximate NNLO~\cite{Kidonakis:2012rm}. The W$+$jets and Z$+$jets cross sections are computed with {\FEWZ} (v3.1)~\cite{Gavin:2010az,Gavin:2012sy} at NNLO, while the diboson cross sections are computed at NLO with {\MCFM} (v6.6)~\cite{Campbell:2011bn}.

For $\ttbar \to \PQb \ell^- \PAGn \PAQb\PQq\PAQq'$ (lepton+jets), $\ttbar \to \PQb \ell^- \PAGn \PAQb \ell^+ \nu$ (dilepton), $\ttbar \to \PQb\PAQq\PQq' \PAQb\PQq\PAQq'$ (all jets), and single top quark processes, six samples with $m_\PQt$ values between 166.5 and 178.5\GeV are generated.

The evaluation of systematic uncertainties related to color reconnection, the modeling of the underlying event, the factorization ($\mu_\mathrm{F}$) and renormalization ($\mu_\mathrm{R}$) scales, and the matching of the parton from the matrix element to parton showers, is based on studies of dedicated samples of simulated events.

A full simulation of the CMS detector based on {\GEANTfour}~\cite{Agostinelli:2002hh} (v9.4p03) is used. Effects of additional overlapping minimum-bias events (pileup) are included in the simulation in such a way that the vertex multiplicity distribution is matched to the data. Single-lepton trigger efficiencies are applied to the simulation to match the trigger selection.

\subsection{Event reconstruction and selection}
\label{subsec:experiment:selection}

Events are reconstructed using a particle-flow (PF) algorithm~\cite{CMS-PAS-PFT-09-001,CMS-PAS-PFT-10-001} that optimally combines the information from all CMS subdetectors to identify and reconstruct individual objects produced in \Pp\Pp\ collisions. The particle candidates include muons, electrons, photons, charged hadrons, and neutral hadrons. Charged particles are required to originate from the primary collision vertex, identified as the reconstructed vertex with the largest value of $\sum \pt^2$ for its associated tracks.
Once isolated muons~\cite{Chatrchyan:2012xi} and electrons~\cite{Khachatryan:2015hwa} are identified and removed from the list of PF particles, charged hadrons are rejected if their tracks do not originate from the primary vertex of the event. Finally, jets are reconstructed from the remaining PF particles using the anti-\kt algorithm~\cite{Cacciari:2008gp} with a distance parameter of 0.5 in the $\eta$--$\phi$ plane. Jet energy corrections are applied to all the jets in data and simulation~\cite{Chatrchyan:2011ds}. The muon \pt scale is corrected to account for possible geometrical effects, such as deformation of tracker geometry still present after implementing the alignment procedure.

The selection criteria are optimized for lepton+jets and dilepton \ttbar events with a \PJGy meson resulting in two additional non-isolated muons.
Lepton+jets events are required to have exactly one isolated lepton with $\pt>26\,(30)\GeV$ and $\abs{\eta}<2.1\,(2.5)$ in the case of the muon (electron). A muon (electron) is considered isolated if the scalar \pt sum of all reconstructed particle candidates (not including the lepton itself) within a cone of size $\Delta R = \sqrt{\smash[b]{(\Delta\eta)^2+(\Delta\phi)^2}} = 0.4$ (0.3) (where the $\phi$ is azimuthal angle in radians) around the lepton direction is less than 12\% (10\%) of the lepton \pt. An event-by-event correction is applied to the scalar sum to take into account possible contributions from pileup events~\cite{Khachatryan:2016mqs}.
Dilepton events are required to have exactly two isolated leptons: at least one isolated lepton defined as above, and either an isolated muon with $\pt>20\GeV$ and $\abs{\eta}<2.4$, or an isolated electron with $\pt>20\GeV$ and $\abs{\eta}<2.5$. In case of a second electron, the isolation threshold is relaxed to less than 15\%. The two leptons are required to be of opposite charge and have an invariant mass above 20\GeV. Pairs with the same flavor and invariant mass between 76 and 106\GeV are rejected to remove poorly reconstructed leptonic Z boson decays.
In addition to these criteria, at least 2 jets with $\pt>40\GeV$ and $\abs{\eta}<2.4$ are required.

Exactly one \PJGy meson candidate, with a mass between 3.0 and 3.2\GeV, is required in the event, reconstructed from two muons of opposite sign, with $\pt>4\GeV$ and $\abs{\eta}<2.4$, that emanate from the same jet. To reduce the combinatorial background, a Kalman vertex fit~\cite{Fruhwirth:1987fm,Fruhwirth:2007hz} with one degree of freedom is performed and the $\chi^2$ of the vertex is required to be less than 5. The significance (\ie, the number of standard deviations) of the distance between the secondary vertex -- formed by the products of the b quark fragmentation -- and the primary vertex of the event is required to be above 20.

These criteria select 666 events in data. The numbers of events expected from the SM processes are evaluated using Monte Carlo (MC) simulation and the results are normalized to their theoretical cross sections. These are noted in Table~\ref{results:tab1}, where a distinction is made between events in which the isolated lepton with the largest \pt is a muon, labeled ``Leading $\mu$'', and events in which the leading isolated lepton is an electron, labeled ``Leading e'', but not between lepton+jets and dileptonic event candidates. The rates predicted by the default simulation are in fair agreement with those observed in data. The event sample is dominated by contributions from lepton+jets and dilepton \ttbar events, with a lesser contribution from single top quark processes.

\begin{table}
\centering
\topcaption{\label{results:tab1}
  Number of selected events from simulation and observed in data. The uncertainties are statistical.
}
\begin{tabular}{lcc}
\hline
\multirow{2}{*}{Process} & \multicolumn{2}{c}{Number of events} \\
 & Leading $\mu$ & Leading $\Pe$ \\
\hline \\[-2.2ex]
$\ttbar \to \PQb \ell^- \PAGn \PAQb\PQq\PAQq'$ & $228.1 \pm 4.0$ & $195.6 \pm 3.7$ \\
$\ttbar \to \PQb \ell^- \PAGn \PAQb \ell^+ \nu$ & $66.3 \pm 1.7$ & $56.9 \pm 1.6$ \\
$\ttbar \to \PQb\PAQq\text{q'} \PAQb\PQq\PAQq'$ & negligible & negligible \\
Single top quark & $39.4 \pm 3.8$ & $30.6 \pm 3.3$ \\
\hline
$\PW\to \ell \nu + \text{jets}$ & $18.3 \pm 3.2$ & $12.1 \pm 2.7$ \\
$\Z / \gamma^* \to \ell^+ \ell^- + \text{jets}$ & $4.5 \pm 0.9$ & $6.3 \pm 1.0$ \\
WW, WZ, ZZ & $1.1 \pm 0.3$ & $1.2 \pm 0.3$ \\
\hline
Predicted yield & $357.7 \pm 6.6$ & $302.7 \pm 5.9$ \\
Data & 355 & 311 \\
\hline
\end{tabular}
\end{table}

\begin{figure}
\centering
\includegraphics[width=0.49\textwidth]{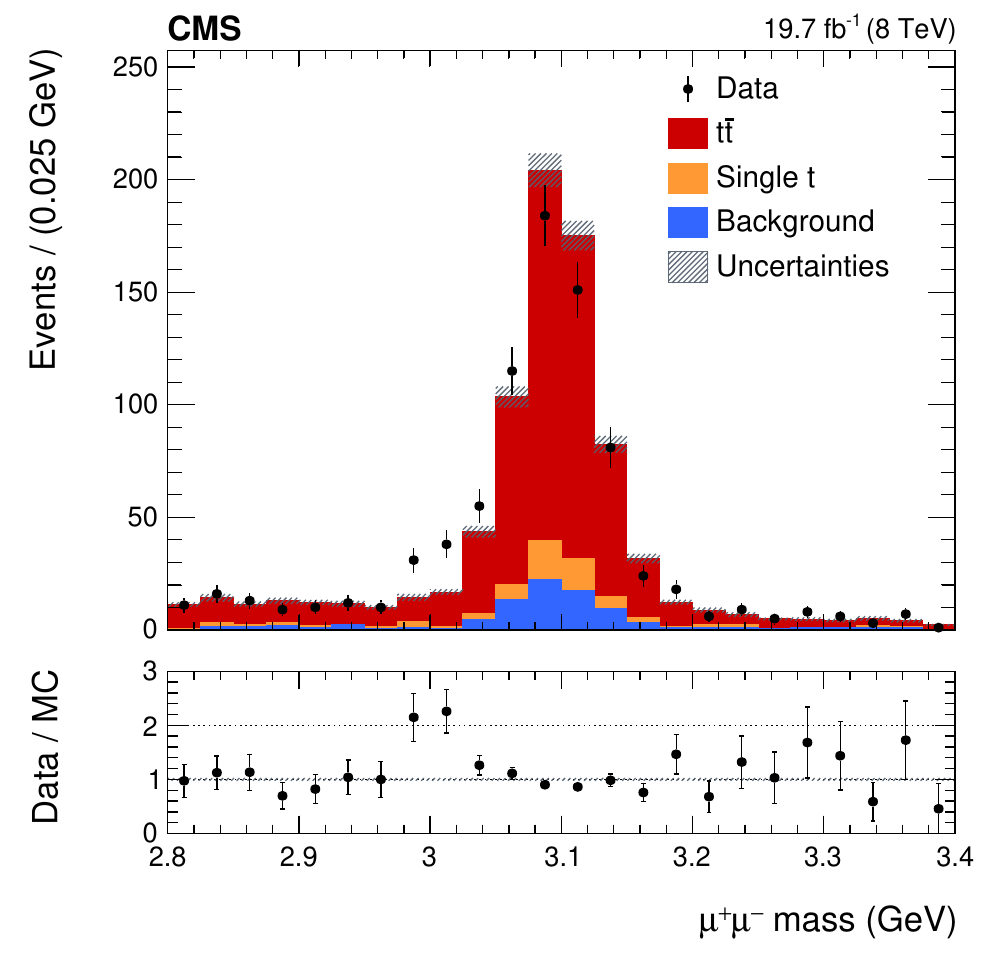}
\includegraphics[width=0.49\textwidth]{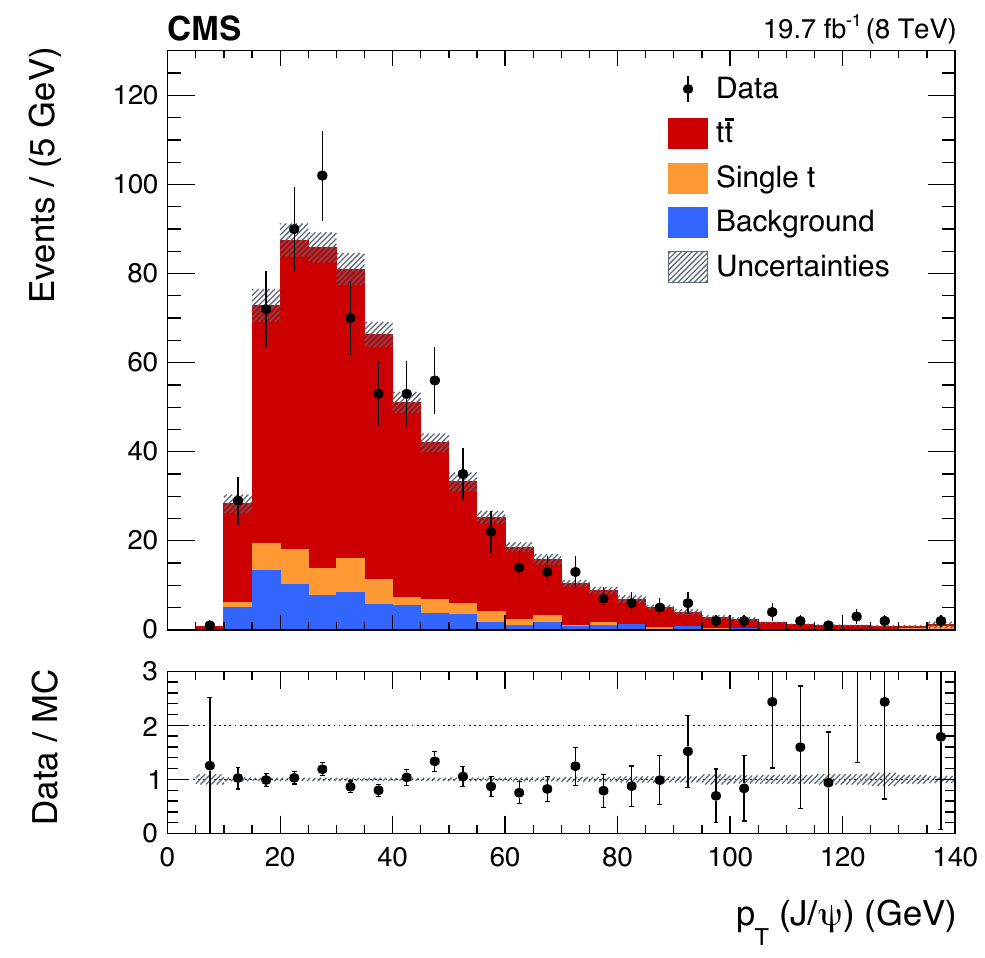}
\caption{Distributions of the dimuon invariant mass between 2.8 and 3.4\GeV~(left) and of the \pt of the \PJGy meson candidate~(right). Processes are normalized to their theoretical cross sections. The simulation assumes a value of $m_\PQt=172.5\GeV$. The lower panel shows the ratio of the number of events observed in data to the number expected from simulation. One point is not visible in the lower panel of  the right plot, as it would require to enlarge the $y$-axis range up to 3.5 units.}
\label{fig:jpsi}
\end{figure}

Figure~\ref{fig:jpsi} shows the dimuon invariant mass spectrum (for a wider mass range than the acceptance window for the \PJGy meson candidates) and the \pt distribution of the \PJGy meson candidates. The simulation used in this figure and the following ones is for $m_\PQt=172.5\GeV$. The ratio in the number of events observed in data to the number expected from simulation is presented in the lower panel. The shaded band includes statistical and systematic uncertainties, which are discussed below, as well as the uncertainty in the integrated luminosity. The number of \PJGy meson candidates is roughly the same in data and simulation. Despite the corrections applied to the muon \pt scale, a worse resolution is observed in data than in simulation. This is caused by final-state radiation emitted by the muons originating from the \PJGy meson decay, which is not included in the simulation~\cite{Khachatryan:2015rra} and which results in a shift of the reconstructed dimuon invariant mass in the simulation to larger values. This effect is included in the systematic uncertainties discussed in Sec.~\ref{subsec:systematics:exp}.

\begin{figure}
\centering
\includegraphics[width=0.49\textwidth]{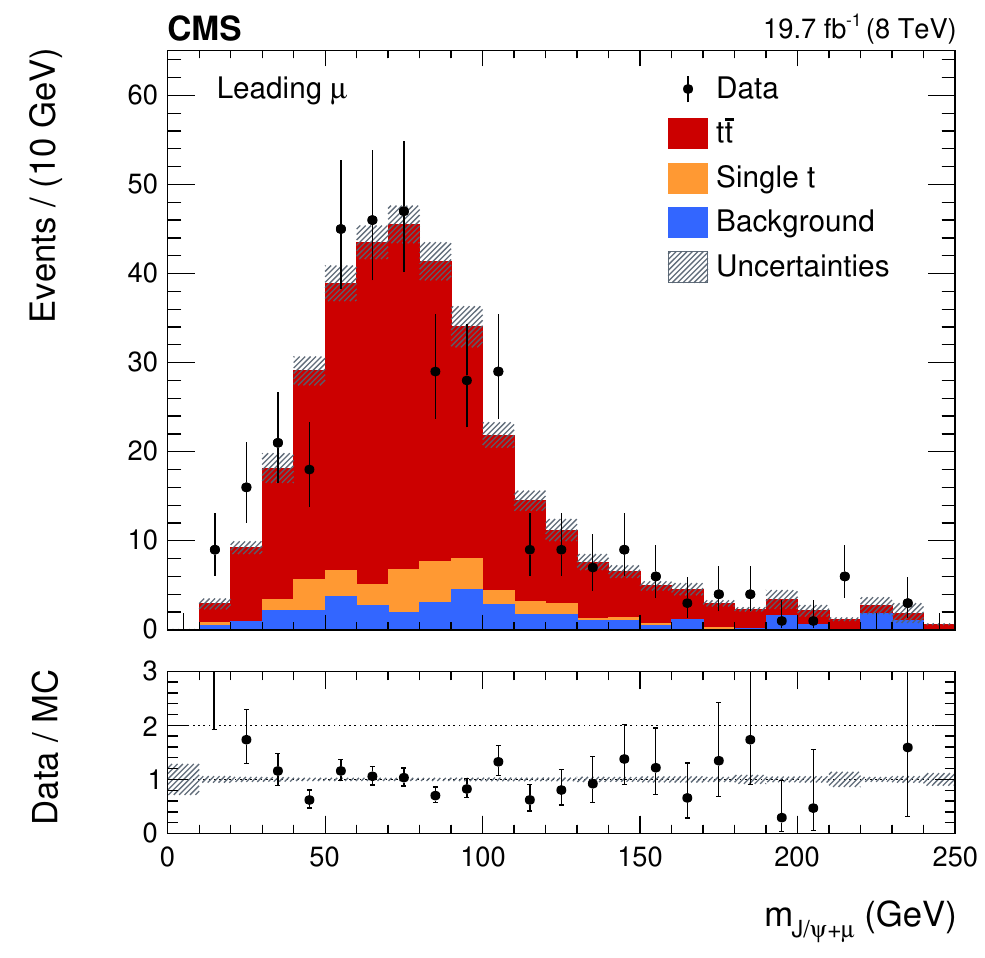}
\includegraphics[width=0.49\textwidth]{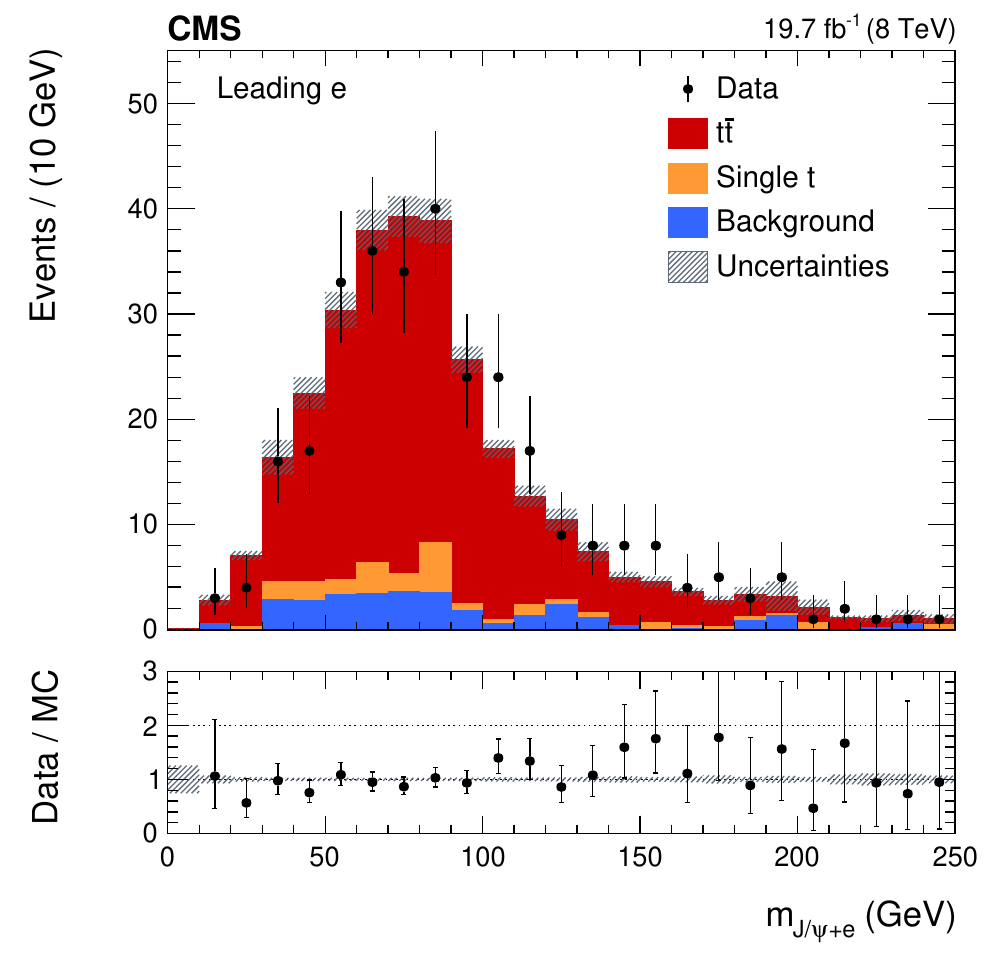}
\caption{Distributions of the invariant mass of the \PJGy meson candidate and the leading lepton combination, in the leading $\mu$ (left) and leading e (right) combinations. Processes are normalized to their theoretical cross sections. The simulation assumes a value of $m_\PQt=172.5\GeV$. The lower panel shows the ratio of the number of events observed in data to the number expected from simulation. Two points are not visible in the lower panel of the left plot and one for the right plot, as it would require to enlarge the $y$-axis range up to 5.5 and 18 units, respectively.}
\label{fig:jpsil}
\end{figure}

The invariant mass, $m_{\PJGy+\ell}$, is computed from the combination of the \PJGy meson candidate and the leading lepton. The distributions are shown in Fig.~\ref{fig:jpsil}.

\section{Extraction of the top quark mass}
\label{sec:measurement}

\subsection{Fitting procedure}
\label{subsec:measurement:procedure}

Since no significant differences are observed between $\PJGy+\mu$ and $\PJGy+\Pe$ events, no further distinction is made on the flavor of the leading lepton.
In associating the leading lepton to a \PJGy meson in a \ttbar event, there are configurations where both particles arise from the same top quark decay chain or from different top quarks (referred to as ``wrong pairings'').
The right and wrong pairings are considered simultaneously in the analysis. While wrong pairings are less sensitive to $m_\PQt$, they remain weakly correlated with it.

The expected $m_{\PJGy+\ell}$ distributions for \ttbar and single top quark processes are simulated for different values of $m_\PQt$. The background contribution is considered to be the same for each $m_\PQt$ value.
The simulated $m_{\PJGy+\ell}$ distributions thus obtained are fitted simultaneously, between 0 and 250\GeV. The signal and background contributions are modelled by the following analytic probability density function:

\begin{multline}
P_\text{sig+bkg}(m_{\PJGy+\ell}) = \alpha \frac{1}{\sigma_g\sqrt{2\pi}}\exp{\left(-\frac{(m_{\PJGy+\ell}-\mu_g)^2}{2\sigma_g^2}\right)} \\
 + (1-\alpha) \frac{\beta_\gamma^{-\gamma_\gamma}}{\Gamma(\gamma_\gamma)}(m_{\PJGy+\ell}-\mu_\gamma)^{\gamma_\gamma-1}\exp{\left(-\frac{m_{\PJGy+\ell}-\mu_\gamma}{\beta_\gamma}\right)}.
   \label{eq:pdf}
\end{multline}

\begin{figure}[htb]
\centering
\includegraphics[width=0.32\textwidth]{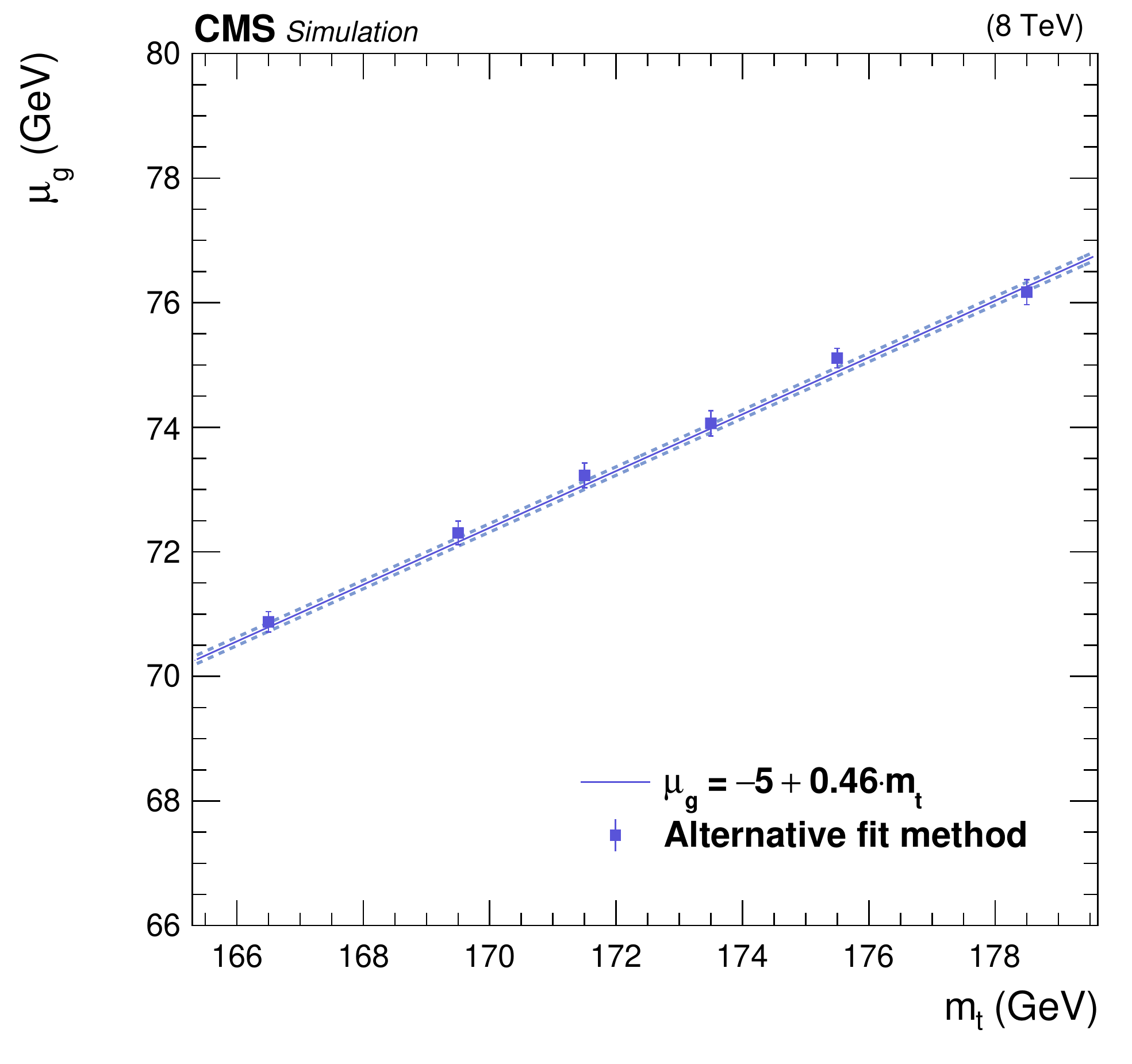} \includegraphics[width=0.32\textwidth]{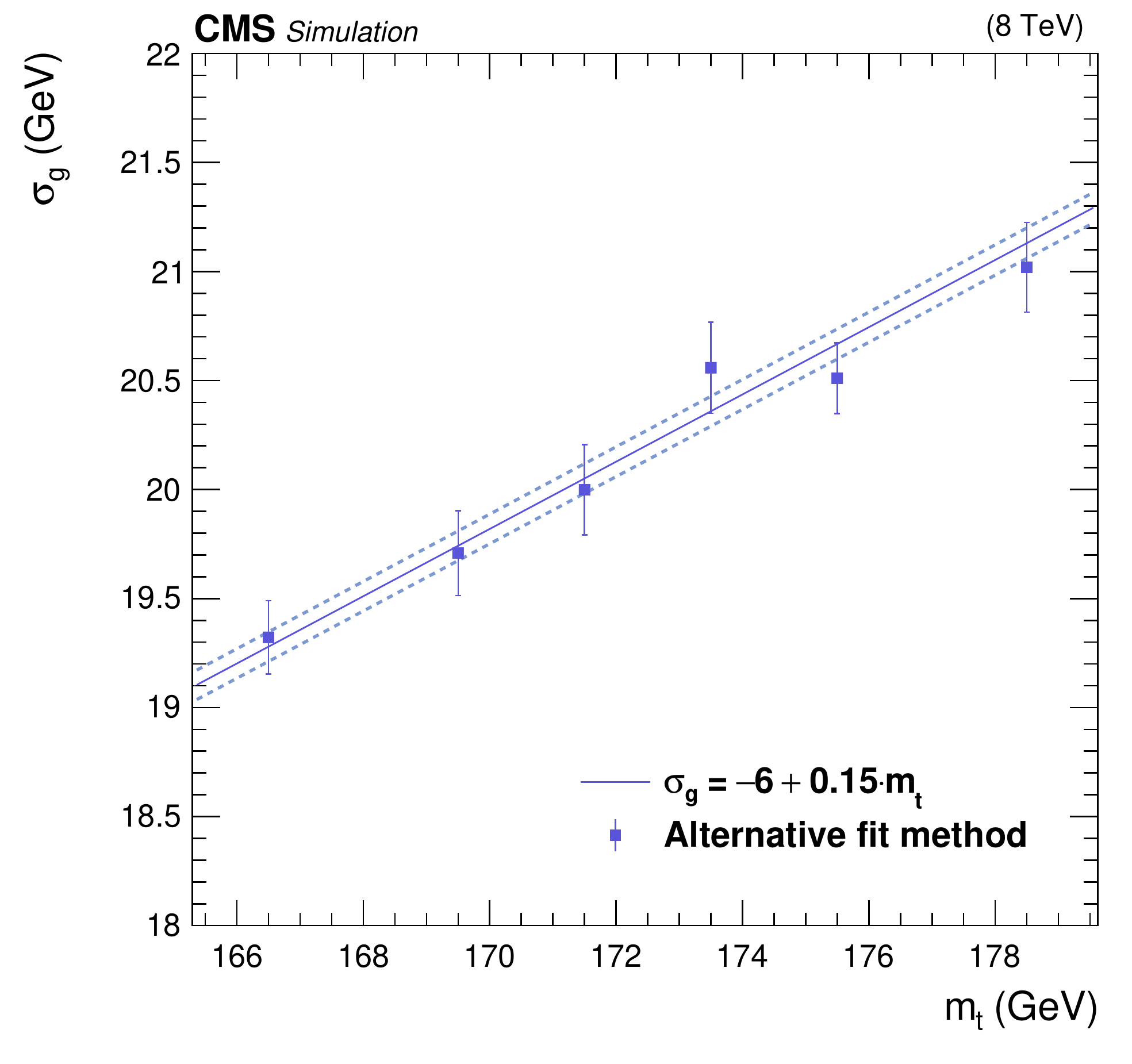}
\includegraphics[width=0.32\textwidth]{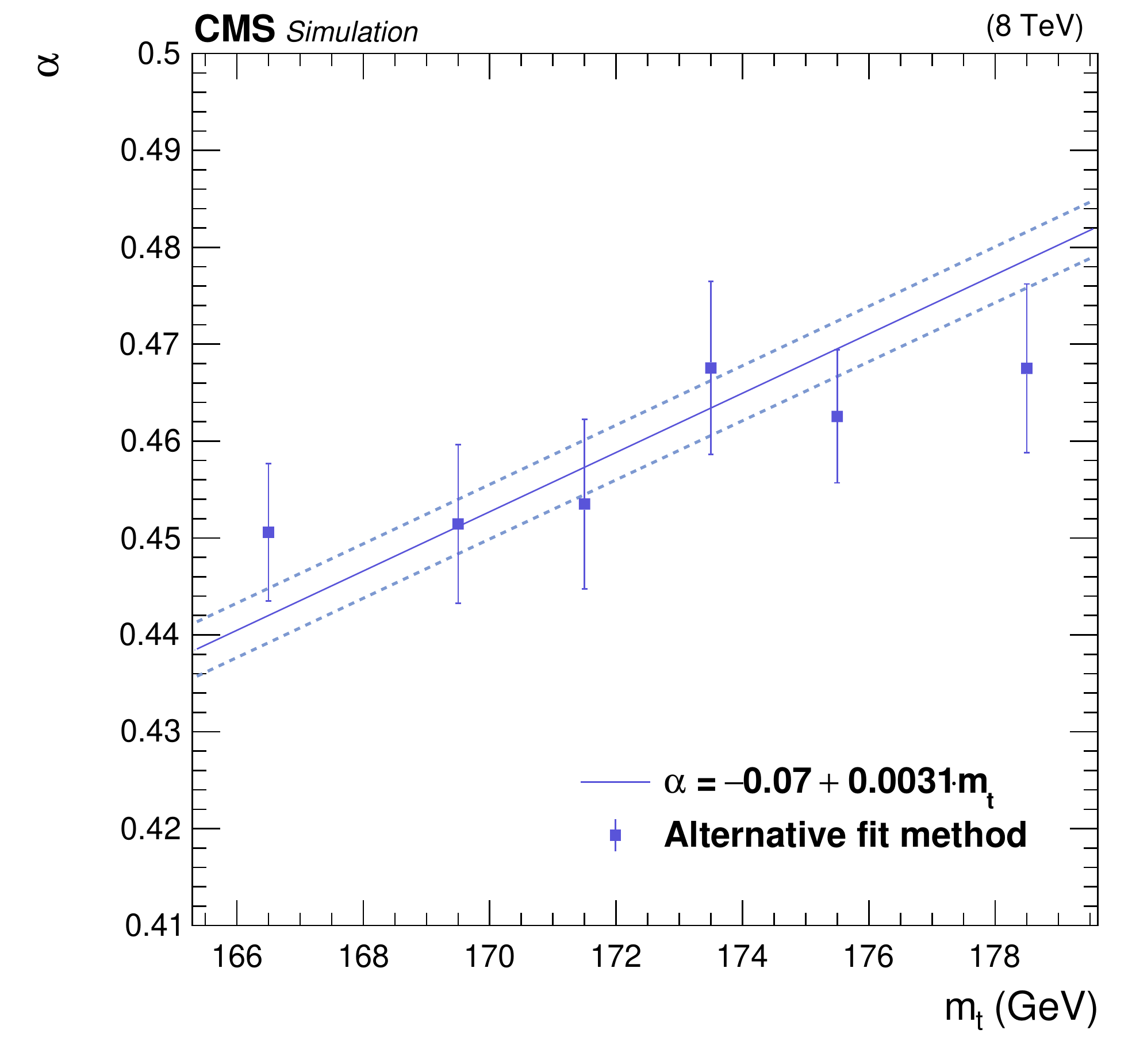} \includegraphics[width=0.32\textwidth]{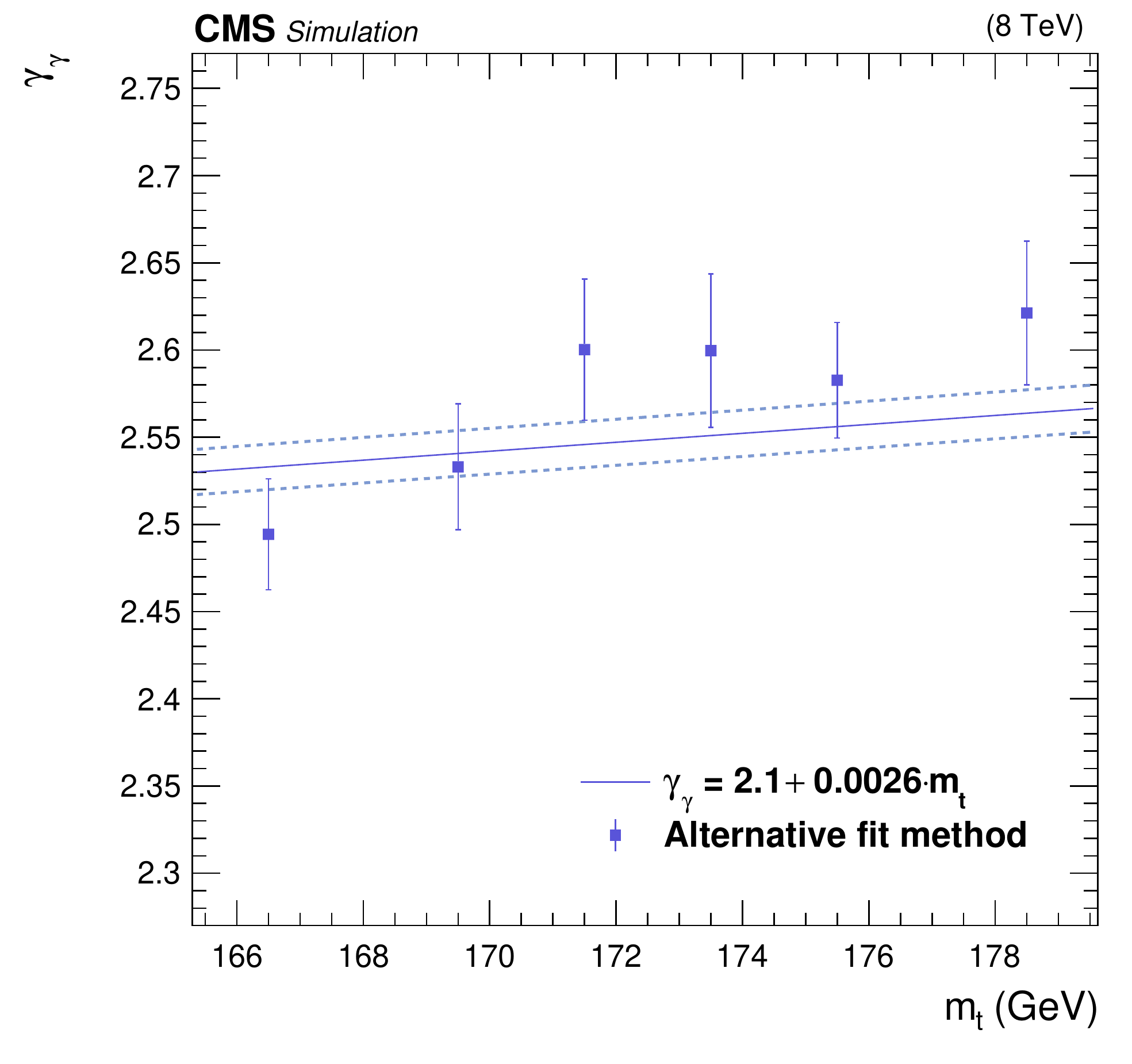} \includegraphics[width=0.32\textwidth]{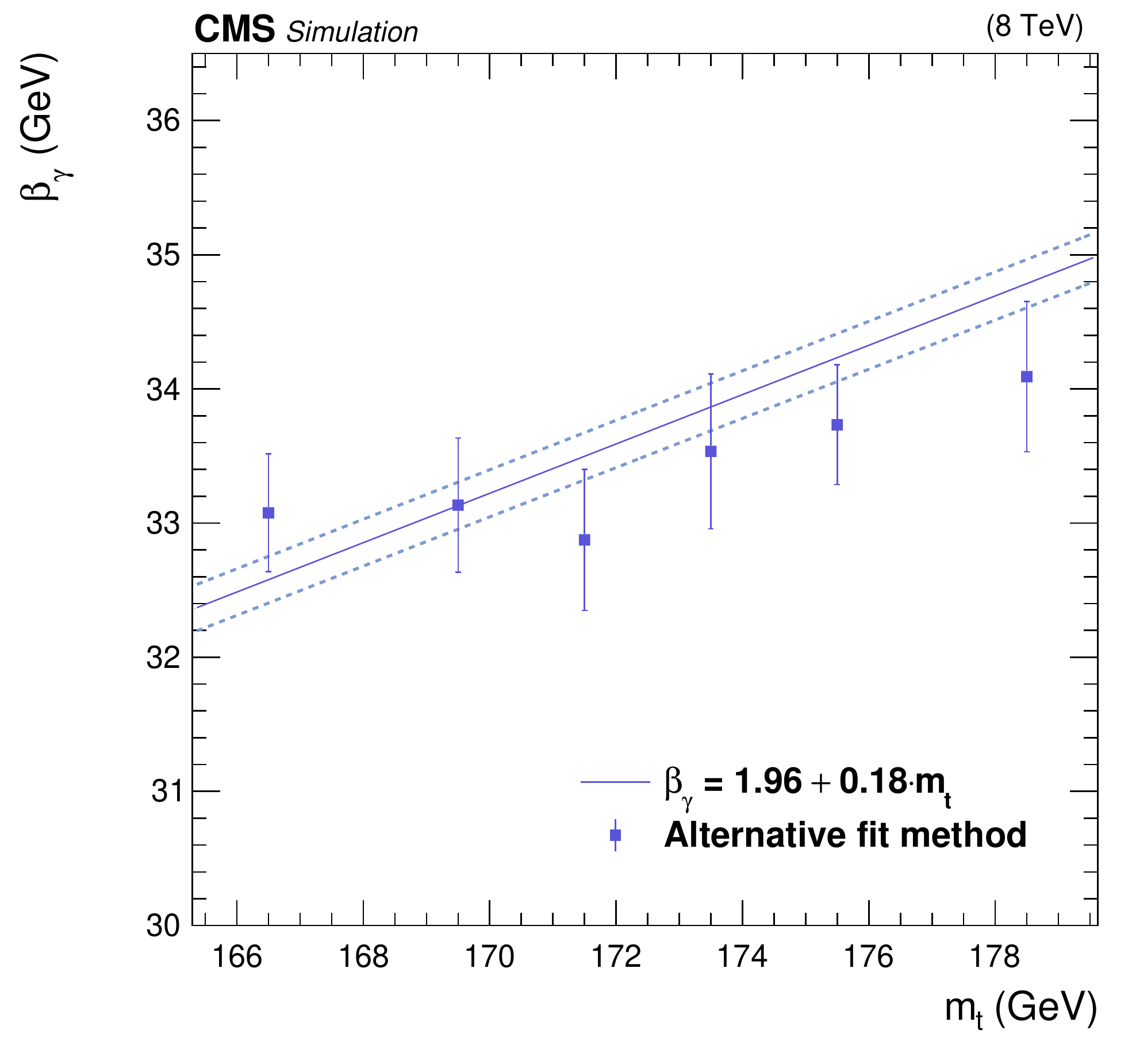} \includegraphics[width=0.32\textwidth]{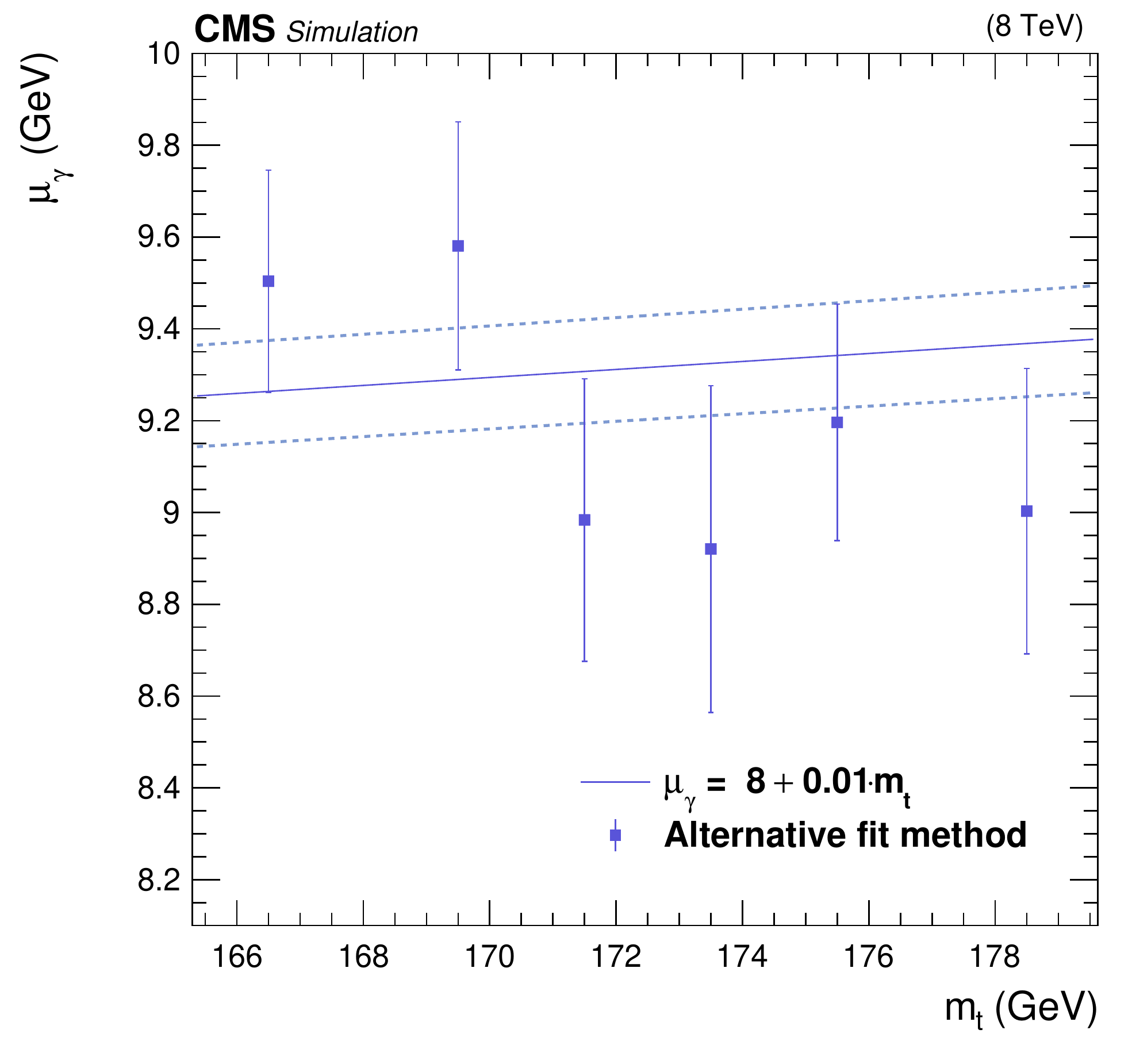} \caption{Mean~(upper left) and standard deviation~(upper middle) of the Gaussian distribution describing the peak of the $m_{\PJGy+\ell}$ distributions, relative contribution of the Gaussian distribution to $P_\text{sig+bkg}$~(upper right), and shape~(lower left), scale~(lower middle), and shift~(lower right) parameters of the gamma distribution, as a function of input $m_\PQt$. The solid lines are the result of the simultaneous fit described in Section~\ref{subsec:measurement:procedure}, while the dashed lines indicate the 68\% confidence level of the fit. The superimposed data points are the result of the alternative fitting method described in Section~\ref{subsec:measurement:xchecks}.}
\label{fig:gaus_par}
\end{figure}

 This function is the sum of a Gaussian distribution (\ie, the first term of the right-hand side in Eq.~(\ref{eq:pdf})) with the free parameters $\mu_g$ (mean) and $\sigma_g$ (standard deviation), which describes mostly the peak in the $m_{\PJGy+\ell}$ distribution, and a gamma distribution (\ie, the whole second term of the right-hand side in Eq.~(\ref{eq:pdf})), whose definition involves the Gamma function $\Gamma$. The gamma distribution has three free parameters: its shape parameter $\gamma_\gamma$, scale parameter $\beta_\gamma$, and shift parameter $\mu_\gamma$. The relative contribution of the Gaussian distribution is described by the parameter $\alpha$.
 Each of the six parameters is implemented as a linear function of $m_\PQt$, taking the form of $c_1+c_2\, m_\PQt$. The $M_{\PJGy+\ell}$ distributions for each of the samples with different values of $m_\PQt$ are simultaneously fitted to obtain the slope and intercept for each of the six parameters. Then, when the $m_{\PJGy+\ell}$ distribution obtained from data is fitted, the linear coefficients $c_1$ and $c_2$ are fixed and $m_\PQt$ becomes the only free parameter of $P_\text{sig+bkg}$.
 Figure~\ref{fig:gaus_par} shows the six parameters of Eq.~(\ref{eq:pdf}) with respect to $m_\PQt$. The two parameters showing the strongest dependence on $m_\PQt$ are $\mu_g$ and $\sigma_g$.

\subsection{Validation of the procedure to extract the top quark mass}
\label{subsec:measurement:xchecks}

Different tests are used to validate the procedure to extract $m_\PQt$.
First, the parameters of $P_\text{sig+bkg}$ are fitted for each of the $m_\PQt$ values independently, without any specific assumption about their dependence on $m_\PQt$. The result, superimposed as the dots in Fig.~\ref{fig:gaus_par}, confirms the assumed linear dependence.
Then the $m_{\PJGy+\ell}$ distribution obtained for $m_\PQt = 172.5\GeV$ is fitted to $P_\text{sig+bkg}$ fixing thereby the dependence of $\mu_g$, $\sigma_g$, $\gamma_\gamma$, $\beta_\gamma$, $\mu_\gamma$, and $\alpha$ on $m_\PQt$, only leaving $m_\PQt$ free. The result is statistically compatible with 172.5\GeV.

The performance of this fitting method is evaluated with pseudo-data experiments.
From $P_\text{sig+bkg}$, described by Eq.~(\ref{eq:pdf}), with $m_\PQt$ fixed at 172.5\GeV, 3\,000 pseudo-data experiments of $N_\text{evt}$ events are drawn, where $N_\text{evt}$ follows a Poisson distribution around the 666 events observed in data.
Each pseudo-data experiment is fitted to $P_\text{sig+bkg}$, with $m_\PQt$ being once again the only free parameter.
The same procedure is reproduced for different $m_\PQt$ values in $P_\text{sig+bkg}$. The residual and the pull, respectively defined as the difference between the fit result and the input value and the difference between the fit result and the input value relative to the fit uncertainty, are computed for each pseudo-data event. The mean and width of the pull and residual distributions obtained for each pseudo-data experiment are rescaled to propagate uncertainties due to the limited numbers of pseudo-data experiments and simulated events.
The means and widths of the pull distributions are found to be constant in $m_\PQt$ and compatible with 0 and 1 within their respective statistical uncertainties. The method to extract $m_\PQt$ from the $m_{\PJGy+\ell}$ distribution can therefore be considered as unbiased.
Each of the six mass points results in a mean and width from the residual distribution, which are interpolated to $m_\PQt = 172.5\GeV$.  The spread of the mean values is used to estimate the systematic uncertainty arising from the size of the simulated event samples (0.22\GeV) and the width values are used to derive the expected statistical uncertainty for the data (2.9\GeV).

\subsection{Modeling heavy-quark fragmentation}
\label{subsec:measurement:fragmentation}

Since this measurement is expected to be particularly sensitive to heavy-quark fragmentation, its corresponding modeling in simulated events is studied in detail.

\begin{figure}
  \centering
  \includegraphics[width=0.55\textwidth]{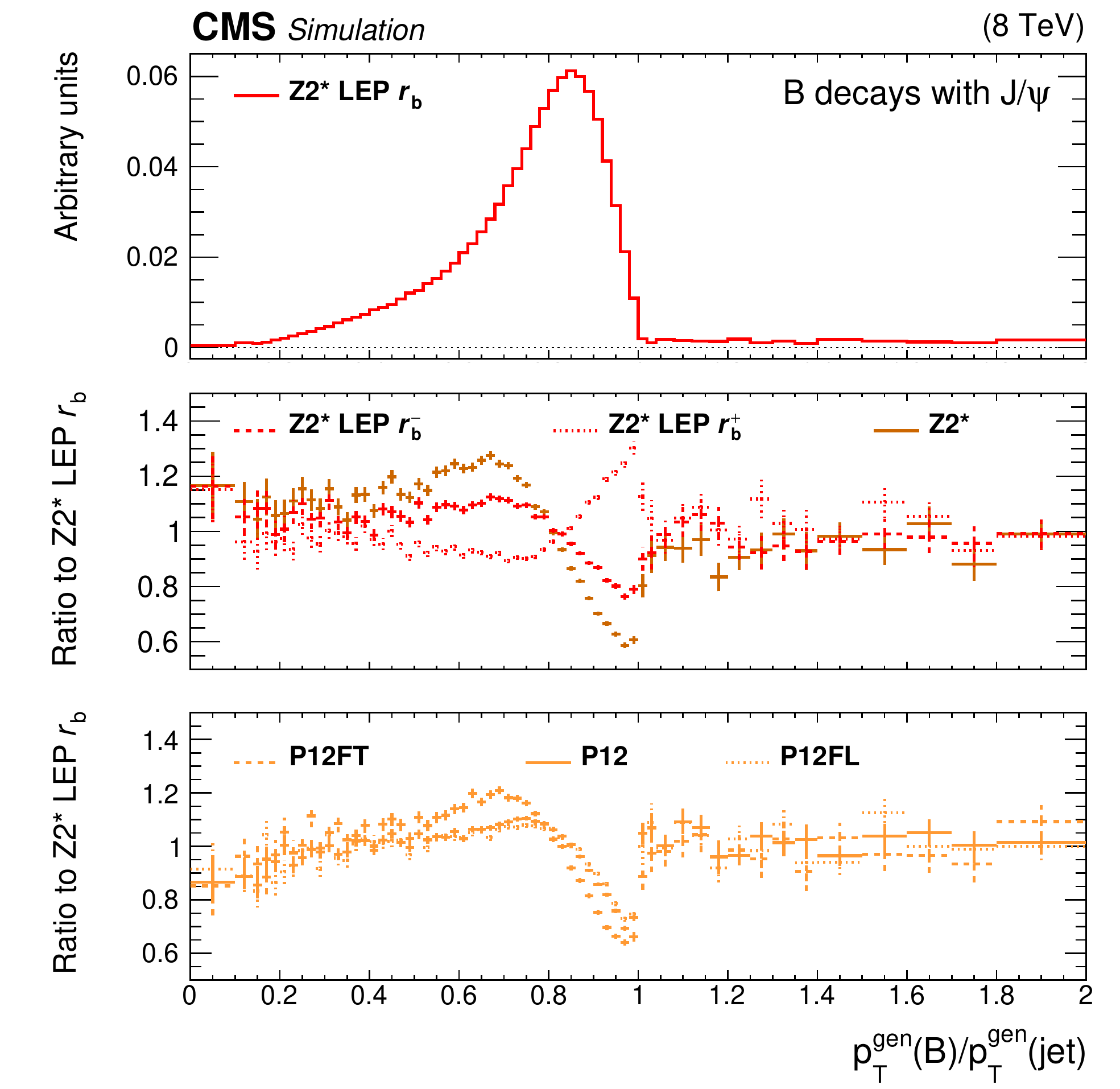}
  \caption{Ratio of the \pt of the b hadrons to the \pt of the matched generator-level jet for the Z2*\;LEP\;$r_\PQb$ tune~(upper), the ratio to Z2*\;LEP\;$r_\PQb$ for the Z2*, Z2*;\text{LEP}\;$r_\PQb^{-}$, and Z2*\;LEP\;$r_\PQb^{+}$ tunes~(middle), and the ratio to Z2*\;LEP\;$r_\PQb$ for the P12, P12FT, and P12FL tunes~(lower). As neutrinos are not clustered within jets, it happens in very rare cases that $\pt^\text{gen}(\PB)>\pt^\text{gen}(\text{jet})$. For this effect to be visible, the horizontal axis range is extended beyond 1 unit.}
\label{fig:pdf_frag}
\end{figure}

\begin{sloppypar}
  The \ttbar simulated event samples in the measurement are generated using the Z2* tune. The \pt distribution of the b hadron at the generator level ($\pt^\text{gen}(\PB)$), relative to that of the jet the hadron is matched to ($\pt^\text{gen}(\text{jet})$), is used to compare the Z2* tune to two alternative tunes and their variants:
\begin{enumerate}
  \item An updated version of the Z2* tune, which better describes fragmentation in $\Pep\Pem$ data, is denoted Z2*\;LEP\;$r_\PQb$~\cite{Khachatryan:2016wqo}. The $r_\PQb$ parameter in the Bowler extension of the fragmentation function~\cite{Bow81} changes from $r_\PQb=1.0$ for Z2* to $0.591$ for Z2*\;LEP\;$r_\PQb$. Values that provide 1 standard deviation changes in the $r_\PQb$ parameter, respectively of $r_\PQb=0.317$ ($\text{Z}2^\ast\;\text{LEP}\;r_\PQb^+$) and~$0.807$ ($\text{Z}2^\ast\;\text{LEP}\;r_\PQb^-$), are also considered;
  \item The Perugia 12 (P12) tune is used along with two variants~\cite{Skands:2010ak} for which the fragmentation process is altered to be harder in the longitudinal (``FL") and transverse (``FT") directions by changing for all quarks the $a$ and $b$ parameters of the Lund fragmentation function~\cite{PR97.1983.31}. The P12 tune is an update of the Perugia 11 tune, used in other analyses, \eg Ref.~\cite{Khachatryan:2016wqo}.
\end{enumerate}
Figure~\ref{fig:pdf_frag} shows the ratio of $\pt^\text{gen}(\PB)/\pt^\text{gen}(\text{jet})$ distribution for the Z2*\;LEP\;$r_\PQb$ tune. For the Z2*, Z2*\;LEP\;$r_\PQb^{-}$, $\text{Z}2^\ast\;\text{LEP}\;r_\PQb^{+}$, P12, P12FT, and P12FL tunes, the ratio to Z2*\;LEP\;$r_\PQb$ is shown. Since this distribution reflects how the \pt of the b quark is transferred to the b hadron, it is a good probe of fragmentation modeling.
\end{sloppypar}

\begin{figure}
  \centering
  \includegraphics[width=0.55\textwidth]{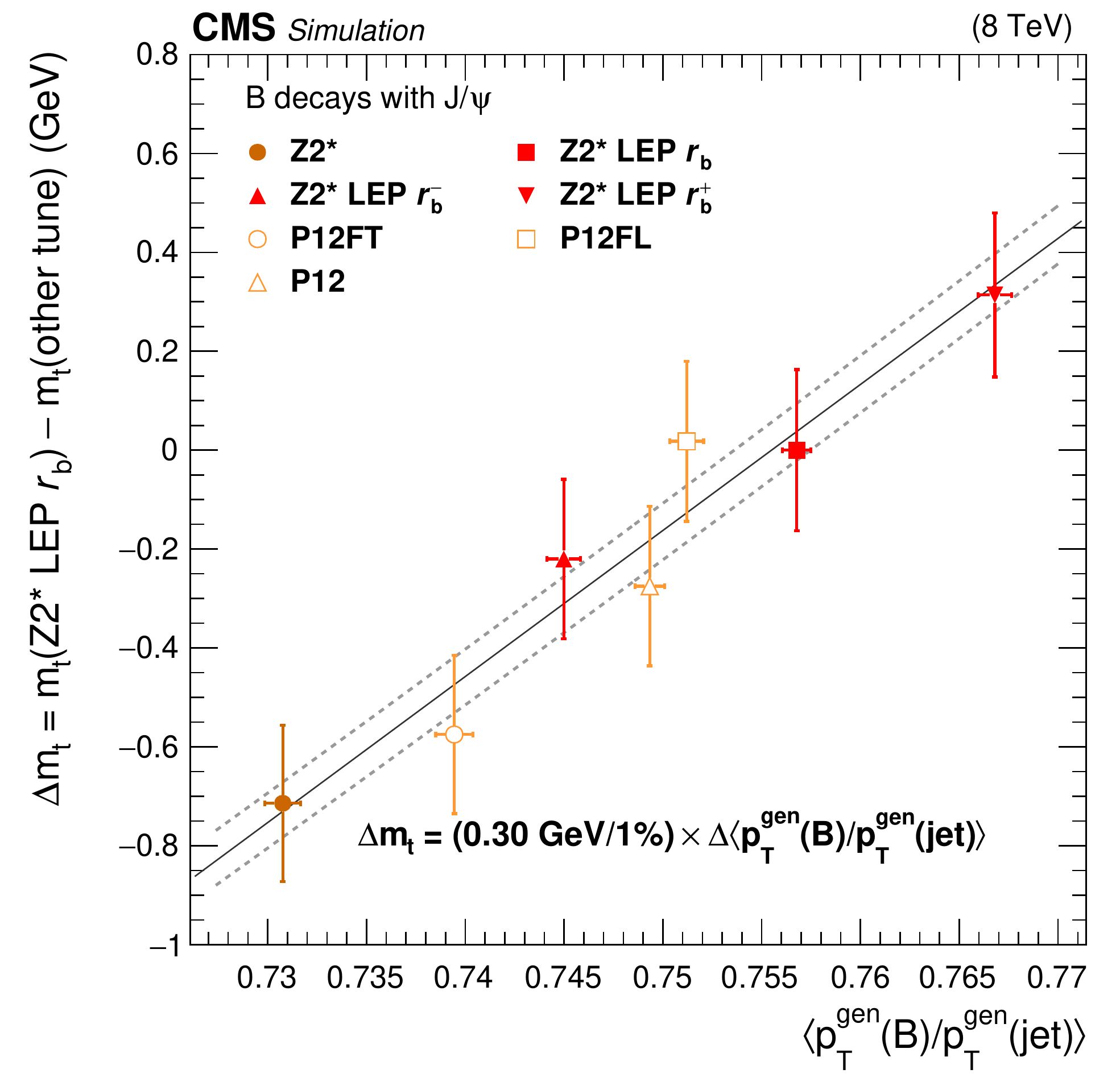}
  \caption{Dependence of the extracted $m_\PQt$ value on the average fragmentation ratio $\langle \pt^\text{gen}(\PB)/\pt^\text{gen}(\text{jet})\rangle$, fitted to a linear function.}
  \label{fig:mtvsxb}
\end{figure}

A reweighting procedure, based on the \pt distribution of the b hadron at generator level relative to that of the jet the hadron belongs to, is applied to the $m_{\PJGy+\ell}$ distribution generated with the Z2* tune at $m_\PQt = 172.5\GeV$. This provides a consistent modeling of the underlying event and color reconnection effects in the Z2* tune, while the description of fragmentation changes. Each reweighted $m_{\PJGy+\ell}$ distribution is then fitted to $P_\text{sig+bkg}$, with $m_\PQt$ being its only free parameter.
Figure~\ref{fig:mtvsxb} shows the dependence of the fitted $m_\PQt$ on the average jet \pt fraction carried by the b hadron in exclusive decays.
In Ref.~\cite{Khachatryan:2016wqo}, it was found that the default Z2* tune is softer than the data for \ttbar events, and the Z2*\;LEP\;$r_\PQb$ tune is a better match to the data.
It appears in Fig.~\ref{fig:mtvsxb} that P12 and Z2* tune families give compatible results within statistical uncertainties.
The Z2*\;LEP\;$r_\PQb$ tune is therefore chosen as the baseline, implying a shift of $-0.71$\GeV to the fit results. The difference between the $m_\PQt$ values obtained for its soft and hard variants is assigned as a systematic uncertainty.

A closure test on the reweighting procedure has been done using simulated event samples generated with the P12 tune family. It validates the strategy of reweighting only the \pt transfer $\pt^\text{gen}(\PB)/\pt^\text{gen}(\text{jet})$.

\subsection{Results}
\label{subsec:measurement:result}

\begin{figure}[h!]
\centering
\includegraphics[width=0.55\textwidth]{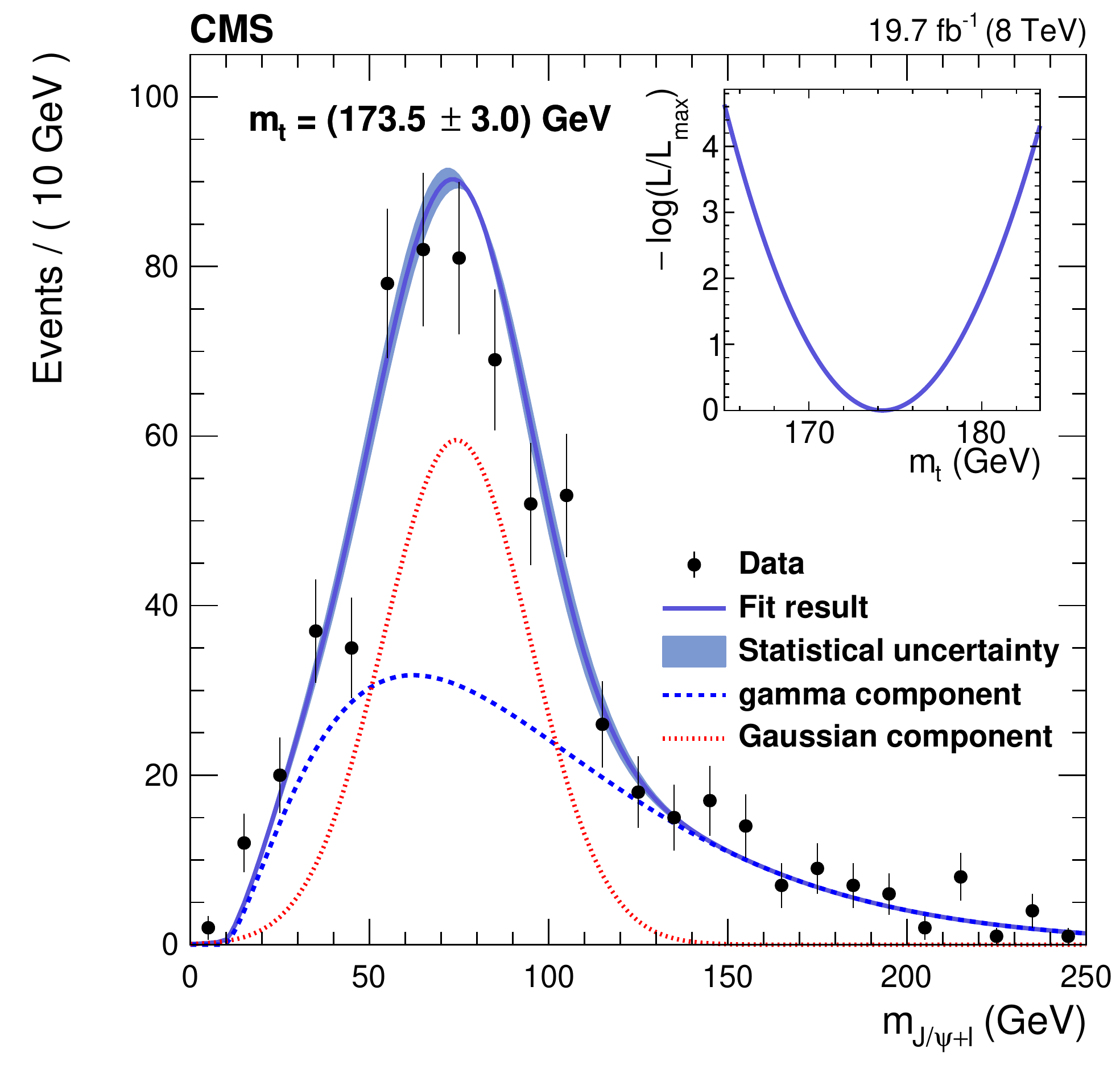}
\caption{Distribution in the invariant mass of the \PJGy meson candidate and the leading lepton combination, fitted to $P_\text{sig+bkg}$ of Eq.~(\ref{eq:pdf}) through the maximization of a likelihood function. The inset shows the negative logarithm of the likelihood function $L$ relative to its maximum $L_\text{max}$ as a function of the only free parameter, which is $m_\PQt$.}
\label{fig:fit_data}
\end{figure}

Figure~\ref{fig:fit_data} shows the $m_{\PJGy+\ell}$ data distribution together with the results of a maximum-likelihood fit to Eq.~(\ref{eq:pdf}). The fit gives a good description of the data, apart from the low mass region, where the missing radiation correction becomes important (see Sec.~\ref{subsec:experiment:selection}). The inset shows the negative logarithm of the likelihood function $L$ relative to its maximum $L_\text{max}$ as a function of $m_\PQt$, which is the only free parameter in the fit. The value of the fitted mass, after implementing the shift of $-0.71$\GeV described in Section~\ref{subsec:measurement:fragmentation}, is 173.5\GeV, with a 68\% confidence level statistical uncertainty of 3.0\GeV.

\section{Systematic uncertainties}
\label{sec:systematics}

The size of each systematic uncertainty is evaluated from its impact on the $m_{\PJGy+\ell}$ shape and its propagation to the fit to extract $m_\PQt$.
For each source of uncertainty, the $m_{\PJGy+\ell}$ distributions are generated for the corresponding variations and then fitted to the nominal parametrization of $P_\text{sig+bkg}$ obtained without variation. A cross-check is performed using pseudo-data experiments. The average shift of $m_\PQt$ with respect to the reference is taken as an estimate of the magnitude of the systematic uncertainty. Both methods are always in good agreement within the statistical uncertainty.

Table~\ref{sys:tab1} summarizes the results obtained for the evaluation of the systematic uncertainties, which are described in detail in Sections~\ref{subsec:systematics:exp} and~\ref{subsec:systematics:th}, and considered as uncorrelated.

\begin{table}
\centering
\topcaption{\label{sys:tab1}
Summary of the impact of systematic uncertainties on the top quark mass according to the contributions from each source.
}
\begin{tabular}{lc}
\hline
Source & Value (\GeVns{}) \\
\hline
\multicolumn{2}{c}{Experimental uncertainties} \\
Limited size of the simulation samples & $\pm$0.22 \\
Muon momentum scale & $\pm0.09$\\
Electron momentum scale & $\pm$0.11 \\
Modeling of the \PJGy meson candidate mass distribution & $+0.09$ \\
Jet energy scale & $<$0.01 \\
Jet energy resolution & $<$0.01 \\
Trigger efficiencies & $\pm$0.02 \\
Pileup & $\pm$0.07 \\
\hline
\multicolumn{2}{c}{Theoretical uncertainties} \\
Background normalization & $\pm$0.01 \\
Matrix-element generator & $-0.37$ \\
Factorization and renormalization scales & $+0.12, -0.46$ \\
Matching of matrix element and parton shower & $+0.12, -0.58$ \\
Top quark transverse momentum & $+0.64$ \\
b quark fragmentation & $\pm$0.30 \\
Underlying event & $\pm$0.13 \\
Modeling of color reconnection & $+0.12$ \\
Parton distribution functions & $+0.39, -0.11$ \\
\hline
Total (in quadrature) &$+0.89, -0.94$ \\
\hline
\end{tabular}
\end{table}

\subsection{Experimental uncertainties}
\label{subsec:systematics:exp}

\paragraph*{Limited size of the simulation samples}
As described in Section~\ref{subsec:measurement:xchecks}, pseudo-data experiments are drawn from $P_\text{sig+bkg}$ for seven different $m_\PQt$ values. The spread of the residual mean is interpreted as the uncertainty due to the finite size of the simulated event samples used for the calibration. No systematic uncertainty stemming from the shape parametrization is added.

\paragraph*{Leading lepton momentum scale}
The average uncertainties in the leading lepton transverse momentum scale are below 0.1\% in the case of muons~\cite{Chatrchyan:2012xi} and 0.3\% in the case of electrons~\cite{Khachatryan:2015hwa}. This uncertainty, given as a function of \pt and $\eta$, is propagated to $m_{\PJGy+\ell}$ and the effect on the $m_\PQt$ fit is evaluated.

\paragraph*{Modeling of the \PJGy meson candidate mass distribution}

Despite the corrections applied to the muon \pt scale, the shape of the \PJGy meson candidate mass distribution observed in data is not exactly reproduced in the simulation, in which final-state radiation from soft muons is not modelled. Conservatively, the full difference is treated as a potential systematic uncertainty. Thus, the $m_{\PJGy+\ell}$ distribution is recomputed for a reweighted \PJGy meson candidate mass, such that the peak position and the width of the simulated distribution are the same as for the data. The uncertainty associated with this effect is computed as the difference between the top quark masses fitted before and after reweighting.

\paragraph*{Jet energy scale and resolution}
In this analysis, the jet energy scale and resolution affect only the event yield.
The effect of the jet energy scale uncertainty is studied by scaling the reconstructed jet energy by a \pt- and $\eta$-dependent scale factor before the event selection is applied.
Similarly, the effect of the jet energy resolution uncertainty is studied by varying the jet energy resolution of the simulated events according to the estimated uncertainty.

\paragraph*{Trigger efficiencies}
As reported in Section~\ref{subsec:experiment:datasim}, the single-lepton trigger efficiencies are applied to simulated events.
A conservative systematic uncertainty of $\pm$3\% is assumed for the trigger efficiencies.
The difference between the top quark masses fitted with upwards and downwards variations is taken as the uncertainty.

\paragraph*{Pileup}
Simulated events are reweighted event by event to reproduce the number of pileup events observed in data. A 5\% variation on the minimum-bias cross section~\cite{Chatrchyan:2012nj} used is propagated to $m_{\PJGy+\ell}$ and $m_\PQt$.

\subsection{Theoretical uncertainties}
\label{subsec:systematics:th}

\paragraph*{Background normalization}
Processes are normalized to their theoretical cross sections. To evaluate the effect of the uncertainties in these cross sections, obtained from scale variations in the theoretical calculation, the main background contributions, \ie $\text{W/Z}+\text{jets}$ and WW/ZZ/WZ, are varied by $\pm$20\% and $\pm$5\%, respectively. Variations in the theoretical cross sections of other processes have negligible impact on the measurement.

\paragraph*{Matrix-element generator}
The {\MADGRAPH} LO matrix-element predictions used for the calibration of the measurement are compared with the NLO {\POWHEG} predictions for $m_\PQt=172.5\GeV$. The difference, propagated to the measured mass, is assigned as a systematic uncertainty.

\paragraph*{Factorization and renormalization scales}
In the signal simulation, $\mu_\text{F}$ and $\mu_\text{R}$ are set to a common value $Q^2=m_\PQt^2+\sum (\pt^\text{parton})^2$, where $\pt^\text{parton}$ is the transverse momentum of the partons. Alternative samples with $Q$ varied by a factor of 0.5 or 2 are used to estimate the effect of the uncertainties in the factorization and renormalization scales.

\paragraph*{Matrix element/parton shower matching threshold}
This matching threshold is a parameter used in the simulation to define the limit at which the generation of extra jets is made by {\PYTHIA} instead of the matrix-element generator {\MADGRAPH}, and therefore controls the hardest initial- and final-state radiation in the event. The effect of the choice of this threshold is evaluated using dedicated samples in which the parameter is changed from the default value of 40\GeV down to 30\GeV and up to 60\GeV, as discussed in Ref.~\cite{Khachatryan:2015mva}.

\paragraph*{Top quark transverse momentum}
Evidence of a mismodeling of the top quark \pt by {\MADGRAPH} has been obtained by the differential cross sections measurements in CMS~\cite{Khachatryan:2015oqa,Khachatryan:2015fwh}.
To quantify the effect of this mismodeling on $m_\PQt$, an event-by-event reweighting is applied to the simulation to reproduce the top quark \pt shape observed in data.
The difference between the top quark masses fitted with and without this reweighting is taken as the uncertainty.

\paragraph*{Fragmentation functions}
The \ttbar simulated event samples used for the measurement are produced with the default Z2* tune, with a correction applied to the fitted result so as to use the Z2*\;LEP\;$r_\PQb$ tune as a baseline. These two tunes are based upon data collected at LEP and elsewhere.
Porting an MC simulation tune from LEP to LHC implies the assumption of the factorization between the perturbative and nonperturbative parts in the shower evolution, which are typically fitted together, and the noncorrelation of these fits with the color structure of the event, which is clearly different in $\Pep\Pem\to\PQb\PAQb$ and $\Pp\Pp\to\ttbar\to\PQb\PWm\PAQb\PWp$ events. These differences are considered to be covered by the underlying event and color reconnection modeling uncertainty. The uncertainty stemming from the modeling of the b hadron decay induces variations of the b hadron relative \pt that are much smaller than the uncertainty in $r_\PQb$ for the Z2*\;LEP\;$r_\PQb$ tune~\cite{Aubert:2002hc}. Thus, only the effect of fragmentation parameters constrained by the LEP data is considered as an additional source of systematic uncertainty, assigning the maximum difference between the $m_\PQt$ values obtained for the Z2*\;LEP\;$r_{\PQb}^\pm$ and Z2*\;LEP\;$r_\PQb$ tunes, shown in Fig.~\ref{fig:mtvsxb}, as the systematic uncertainty stemming from the fragmentation modeling. The size of the uncertainty is found to be comparable to the one estimated in a different way in Ref.~\cite{Kharchilava:1999yj}.

\paragraph*{Hadronization modeling}
A generator-level study using {\SHERPA} (v2.1.0)~\cite{Gleisberg:2008ta} has been carried out in the context of Ref.~\cite{Khachatryan:2016wqo}. The {\SHERPA} generator allows us to use the same \pt-ordered shower model (\textsc{csshower++}~\cite{Schumann:2007mg}), while interfacing with two alternative hadronization/fragmentation models. The difference between the cluster and the string models on the $m_{\PJGy+\ell}$ shape is much smaller than the difference between the Z2*\;LEP\;$r_{\PQb}^{+}$ and Z2*\;LEP\;$r_{\PQb}^{-}$ tunes. Thus, only the difference between the two fragmentation tunes is considered as a source of systematic uncertainty and no extra uncertainty stemming from the hadronization model is assigned.

\paragraph*{Underlying event and color reconnection modeling}
These effects are evaluated using variations of the Perugia 12 (P12) underlying event tune~\cite{Skands:2010ak}. Two variations (``ueHi" and ``ueLo") are compared to the nominal P12 tune to evaluate the effect of the underlying event in the measurement. The nominal P12 tune is taken here as the reference as it contains not only a dedicated parametrization of the fragmentation function, but also different parametrizations for the hadron multiplicities.
The difference between the nominal P12 tune and a separate variation where color reconnection effects are smaller (``crLo") is assigned as the systematic uncertainty due to this effect.

\paragraph*{Parton distribution functions}
\begin{sloppypar}
As stated in Section~\ref{subsec:experiment:datasim}, the default PDF tune is CTEQ6L1 for \ttbar simulated events in this analysis. The $m_\PQt$ value fitted in this tune is compared to the one obtained for the CT14 NLO~\cite{Dulat:2015mca}, MMHT2014 NCL 68CL~\cite{Harland-Lang:2014zoa}, and NNPDF30 NLO AS0118~\cite{Ball:2014uwa} tunes, applying the PDF4LHC recommendations~\cite{Alekhin:2011sk,Botje:2011sn}. Diagonalized uncertainty sources of each PDF set are used to derive event-by-event weights, which are then applied to obtain a variation of the $M_{\PJGy+\ell}$ shape. The maximal difference with respect to the nominal $M_{\PJGy+\ell}$ shape is quoted as the systematic uncertainty.
\end{sloppypar}

\section{Summary}
\label{sec:conclusion}

The first measurement of the mass of the top quark is presented in the decay channel $\PQt\to(\PW\to\ell\nu)\,(\PQb\to\PJGy+\mathrm{X}\to\PGmp\PGmm+\mathrm{X})$.
An event selection is implemented in proton-proton collisions recorded with the CMS detector at $\sqrt{s}=8\TeV$, to obtain a sample of high purity leptonically-decaying top quarks in \ttbar and single top quark production events containing one \PJGy meson candidate that decays into an oppositely-charged muon pair.
The data correspond to an integrated luminosity of 19.7\fbinv.
There are 355 events observed with a muon and 311 with an electron as leading isolated lepton, in agreement with expectations from simulation.

The top quark mass is extracted from an unbinned maximum-likelihood fit to the invariant mass of the leading lepton and \PJGy meson candidate. The resulting $m_\PQt$ measurement is 173.5\GeV, with a statistical uncertainty of 3.0\GeV and a systematic uncertainty of 0.9\GeV.
This is the first time that this method has been applied to a physics analysis and the systematic uncertainty is of the same order of magnitude as that estimated in Ref.~\cite{Kharchilava:1999yj}.
Even though the results are statistically limited, the dominant systematic uncertainties are different from those of the most precise direct reconstruction methods. As the sensitivity to jet-related uncertainties is negligible, this allows the possibility to contribute significantly in combination with other $m_\PQt$ measurements. Furthermore, with the larger data set expected in the next runs of the LHC, the method described in this paper will provide a result which will be more competitive with those obtained from the conventional reconstruction techniques.

\begin{acknowledgments}
\hyphenation{Rachada-pisek}
We congratulate our colleagues in the CERN accelerator departments for the excellent performance of the LHC and thank the technical and administrative staffs at CERN and at other CMS institutes for their contributions to the success of the CMS effort. In addition, we gratefully acknowledge the computing centers and personnel of the Worldwide LHC Computing Grid for delivering so effectively the computing infrastructure essential to our analyses. Finally, we acknowledge the enduring support for the construction and operation of the LHC and the CMS detector provided by the following funding agencies: BMWFW and FWF (Austria); FNRS and FWO (Belgium); CNPq, CAPES, FAPERJ, and FAPESP (Brazil); MES (Bulgaria); CERN; CAS, MoST, and NSFC (China); COLCIENCIAS (Colombia); MSES and CSF (Croatia); RPF (Cyprus); SENESCYT (Ecuador); MoER, ERC IUT and ERDF (Estonia); Academy of Finland, MEC, and HIP (Finland); CEA and CNRS/IN2P3 (France); BMBF, DFG, and HGF (Germany); GSRT (Greece); OTKA and NIH (Hungary); DAE and DST (India); IPM (Iran); SFI (Ireland); INFN (Italy); MSIP and NRF (Republic of Korea); LAS (Lithuania); MOE and UM (Malaysia); BUAP, CINVESTAV, CONACYT, LNS, SEP, and UASLP-FAI (Mexico); MBIE (New Zealand); PAEC (Pakistan); MSHE and NSC (Poland); FCT (Portugal); JINR (Dubna); MON, RosAtom, RAS and RFBR (Russia); MESTD (Serbia); SEIDI and CPAN (Spain); Swiss Funding Agencies (Switzerland); MST (Taipei); ThEPCenter, IPST, STAR and NSTDA (Thailand); TUBITAK and TAEK (Turkey); NASU and SFFR (Ukraine); STFC (United Kingdom); DOE and NSF (USA).

Individuals have received support from the Marie-Curie program and the European Research Council and EPLANET (European Union); the Leventis Foundation; the A. P. Sloan Foundation; the Alexander von Humboldt Foundation; the Belgian Federal Science Policy Office; the Fonds pour la Formation \`a la Recherche dans l'Industrie et dans l'Agriculture (FRIA-Belgium); the Agentschap voor Innovatie door Wetenschap en Technologie (IWT-Belgium); the Ministry of Education, Youth and Sports (MEYS) of the Czech Republic; the Council of Science and Industrial Research, India; the HOMING PLUS program of the Foundation for Polish Science, cofinanced from European Union, Regional Development Fund, the Mobility Plus program of the Ministry of Science and Higher Education, the National Science Center (Poland), contracts Harmonia 2014/14/M/ST2/00428, Opus 2013/11/B/ST2/04202, 2014/13/B/ST2/02543 and 2014/15/B/ST2/03998, Sonata-bis 2012/07/E/ST2/01406; the Thalis and Aristeia programs cofinanced by EU-ESF and the Greek NSRF; the National Priorities Research Program by Qatar National Research Fund; the Programa Clar\'in-COFUND del Principado de Asturias; the Rachadapisek Sompot Fund for Postdoctoral Fellowship, Chulalongkorn University and the Chulalongkorn Academic into Its 2nd Century Project Advancement Project (Thailand); and the Welch Foundation, contract C-1845.
\end{acknowledgments}

\bibliography{auto_generated}

\cleardoublepage \appendix\section{The CMS Collaboration \label{app:collab}}\begin{sloppypar}\hyphenpenalty=5000\widowpenalty=500\clubpenalty=5000\textbf{Yerevan Physics Institute,  Yerevan,  Armenia}\\*[0pt]
V.~Khachatryan, A.M.~Sirunyan, A.~Tumasyan
\vskip\cmsinstskip
\textbf{Institut f\"{u}r Hochenergiephysik der OeAW,  Wien,  Austria}\\*[0pt]
W.~Adam, E.~Asilar, T.~Bergauer, J.~Brandstetter, E.~Brondolin, M.~Dragicevic, J.~Er\"{o}, M.~Flechl, M.~Friedl, R.~Fr\"{u}hwirth\cmsAuthorMark{1}, V.M.~Ghete, C.~Hartl, N.~H\"{o}rmann, J.~Hrubec, M.~Jeitler\cmsAuthorMark{1}, A.~K\"{o}nig, I.~Kr\"{a}tschmer, D.~Liko, T.~Matsushita, I.~Mikulec, D.~Rabady, N.~Rad, B.~Rahbaran, H.~Rohringer, J.~Schieck\cmsAuthorMark{1}, J.~Strauss, W.~Treberer-Treberspurg, W.~Waltenberger, C.-E.~Wulz\cmsAuthorMark{1}
% ...
\vskip\cmsinstskip
% ...
1:~~Also at Vienna University of Technology, Vienna, Austria\\
% ...

\end{sloppypar}
\end{document}